\documentclass[twocolumn,showpacs]{revtex4}  % Use  for  single column
\usepackage{graphicx}%
\newcommand{\figwidth}{3.5in} % Use for single column output
\begin{document}
%\twocolumn[\hsize\textwidth\columnwidth\hsize\csname @twocolumnfalse\endcsname
% \draft comtempremand makes pacs numbers print
%\draft
\title{Scaling  of thermal  conductivity  of  helium confined  in
pores}
% repeat the \author\address pair as needed
\author{Kwangsik       Nho       and      Efstratios       Manousakis}
\affiliation{Department   of  Physics   and  MARTECH,   Florida  State
University, Tallahassee, Florida 32306} \date{\today}
\begin{abstract}
We have studied the thermal  conductivity of confined superfluids on a
bar-like geometry. We use the planar magnet lattice model on a lattice
$H\times  H\times L$ with  $L \gg  H$. We  have applied  open boundary
conditions on  the bar sides  (the confined directions of  length $H$)
and periodic along the long direction.   We have adopted a hybrid Monte
Carlo algorithm to efficiently deal with the critical slowing down and
in  order  to  solve  the  dynamical  equations of  motion  we  use  a
discretization  technique  which  introduces  errors  only  $O((\delta
t)^6)$ in the time step $\delta t$.  Our results demonstrate the validity
of scaling using known values of the critical exponents and
we  obtained the  scaling function of the thermal resistivity.  
We find that our results for the
thermal resistivity  scaling function are in very  good agreement with
the  available experimental  results for  pores using  the temperature
scale and thermal resistivity scale as free fitting parameters.
\end{abstract}

% insert suggested PACS numbers in braces on next line
\maketitle

%]                  
\section{Introduction}
\label{sec0}
The  superfluid   transition  of  liquid  $^{4}He$   offers  a  unique
opportunity for  testing the finite-size scaling theory  of static and
dynamic  critical phenomena.  Recently,  a sophisticated  experimental
study  was  carried out  in  microgravity  environment, the  so-called
confined helium experiment (CHeX). Lipa et al.\cite{chex} measured the
specific heat of  helium confined in a parallel  plate geometry with a
spectacular   nanokelvin  resolution,  thus,   providing  experimental
results within  a few nanokelvin of $T_{\lambda}$.   When the critical
temperature is  approached, the bulk  correlation length $\xi$  of the
fluid  can  become  of  the  order  of  the  confining  length.   CHeX
approached so  close to the  lambda point that the  correlation length
became macroscopic  in size.  In this  case the whole fluid  acts in a
correlated way and  this changes the value of  global properties, such
as the  specific heat,  relative to their  bulk values. In  a parallel
approach, Mehta  and Gasparini\cite{metha1,metha2} have  also reported
earth-bound  measurements  on   samples  with  smaller  plate  spacing
$L$. The size of $L$ used is these measurements is smaller so that the
results   are  not   significantly   influenced  by   the  change   in
$T_{\lambda}$  between the  top  and  bottom of  the  film because  of
hydrostatic  pressure  difference  which  exists due  to  the  earth's
gravitational field.

The   finite-size   scaling    (FSS)   theory\cite{finite}   and   the
renormalization-group theory (RGT)\cite{rgt} were expected to describe
the behavior of  the system at temperature close  to $T_{\lambda}$.  A
testable implication of  this theory is that very  close to the lambda
point, in  a confined system  with a confining  length of size  $H$, a
dimensionless quantity or the ratio  of two quantities having the same
dimensions, is only a function  of the ratio of $\xi/H$. Therefore the
values of a  given observable $O(t,H)$, for various  values of $H$ and
of the reduced temperature  $t=|1-T/T_{\lambda}|$, divided by its bulk
value of  $O(t,H=\infty)$ should  be a dimensionless  scaling function
$f(x)$,  where $x=\xi(t)/H$. The  results of  CHeX were  in remarkable
agreement  with   predictions  which  were  available   prior  to  the
experiment    based    on     scaling    functions    obtained    from
renormalization-group   theory\cite{the1}   and   those  obtained   by
combining   FSS   and    the   results   obtained   from   large-scale
simulations\cite{sm95}.

A second  equally important step toward understanding  the FSS theory
is  to  study  dynamical  and  transport properties  near  a  critical
point. A well-suited  candidate problem for this study  is the thermal
conductivity   $\lambda$  of   $^{4}He$   near  $T_{\lambda}$.    When
$T_{\lambda}$ is  approached from  above, the thermal  conductivity of
the  fluid diverges\cite{ahlers1,kerrisk}.   The  precise behavior  of
bulk $\lambda$ as  a function of $t$ was studied  in great detail both
experimentally\cite{tam,dingus,tam2} and theoretically\cite{hal}.

There  are several  recent theoretical  studies of  dynamical critical
phenomena and dynamical scaling.  Koch, Dohm, and Stauffer\cite{koch1}
presented field-theoretical  and numerical studies of  the validity of
dynamic  finite-size  scaling   for  relaxational  dynamics  in  cubic
geometry   with   periodic  boundary   conditions   above  and   below
$T_{c}$. Quantitative  agreement between  theory and Monte  Carlo data
was  obtained  by them.  Koch  and  Dohm\cite{koch2}  have provided  a
prediction  for  the  dynamic  finite-size scaling  function  for  the
effective   diffusion  constant   of  model   $C$  of   Hohenberg  and
Halperin\cite{hohenberg}.     Bhattacharjee\cite{bha}    derived    an
approximate form of the  scaling function for the thermal conductivity
using  a  decoupled-mode  approximation  and  model  $E$.   Krech  and
Landau\cite{krech}  calculated   the  transport  coefficient   of  the
out-of-plane magnetization  component at the critical  point, which is
related to the thermal conductivity of liquid $^{4}He$ using Monte 
Carlo spin
dynamics simulations of the $XY$ model in three dimensions on a simple
cubic lattice with periodic  boundary conditions.  They determined the
critical exponent characterizing the thermal conductivity.

Accurate  experimental studies  have  been carried  out  not only  for
dynamic  bulk  phenomena with  improved  resolution  but also  dynamic
properties  {\it  in  confined  geometries}  deeply  in  the  critical
region\cite{lipa,ahlers2}.      Rather      recently,     Kahn     and
Ahlers\cite{kahn}  measured  the   thermal  conductivity  of  
liquid $^{4}He$
confined in  a glass  capillary array of  thickness 3 $mm$  with holes
$2\mu  m $  in diameter.   Their  results show  that long  cylindrical
samples  have a transition  from three-dimensional  to one-dimensional
behavior  and there  is  no phase  transition  in the  one-dimensional
system. However,  as measurements over  a wide range  of hole-diameter
are  required in  order to  test  the finite-size  scaling theory  for
transport    properties,     further    experimental    studies    are
planned\cite{best}  in order  to reveal  dynamical exponents  near the
critical point  and to study  the finite-size scaling behavior  of the
thermal  conductivity  in such  confining  geometries.   To avoid  the
limitations  imposed   by  the  earth's   gravity,  this  experimental
effort\cite{best} will be carried out under microgravity conditions on
the  Low  Temperature  Microgravity  Facility on  International  Space
Station.

In this paper  we wish to study the  thermal conductivity $\lambda$ of
confined helium and to  calculate the scaling function associated with
$\lambda$ for a fixed  geometry.  Since there are already experimental
results\cite{kahn}  for  the scaling  function  of  $\lambda$ for  the
pore-like geometry, in this paper  we will focus our attention to this
geometry  because we  hope to  compare with  the experiment.   We will
examine the FSS theory for the thermal conductivity of helium confined
in  a bar-like  geometry i.e.,  on an  $H^{2} \times  L$  lattice with
$L>>H$.   This confining  geometry  is  similar to  that  of Kahn  and
Ahlers\cite{kahn}  because two  of the  dimensions of  a pore  used in
their  experimental  studies are  confining  as  is  the case  of  the
bar-like geometry.   We will consider  the limit in which  our results
are  independent  of the  bar  length  $L$.   We will  apply  periodic
boundary  conditions  (BC)  in  the  $L$ direction  because  these  BC
approach the bulk limit faster.  In the other two directions which are
kept finite we  will apply open boundary conditions.   We will use the
dynamics of  planar-magnet model and  Monte Carlo simulation  to study
$\lambda(t,H)$.  We find that $\lambda(t,H) H^{-\pi/\nu}$ plotted as a
function  of $x=tH^{1\over \nu}$  fall on  the same  curve for  a wide
range of  values of $H$ and $t$,  using the known values  of $\nu$ and
$\pi$.  This demonstrates that finite-size is also valid for dynamical
critical properties.  In addition we obtain the scaling function which
fits  very well the  experimental data  of Kahn  and Ahlers\cite{kahn}
using the scale  of temperature and the thermal  conductivity scale as
free parameters.

\section{The Method}

We will first briefly describe the model and the numerical method used
and show  how the thermal conductivity  is computed in  our model.  To
describe the dynamics  of a superfluid, we will  use the planar magnet
model which  is classified as model $F$  (or $E$ in the  absence of an
external field) by  Hohenberg and Halperin\cite{hohenberg}.  Matsubara
and Matsuda\cite{mat} has proposed model $F$ to explain the properties
of liquid $^{4}He$. The problem of hard core bosons can be described
by a lattice gas model which can be mapped to the quantum antiferromagnet
in which the superfluid order parameter corresponds
to $S_x-i S_y$ while the density of the boson system corresponds
to $1/2-S_z$. In order to study equilibrium critical properties of a 
superfluid one uses  the $XY$ model\cite{sm95,sm} because the  planar magnet
model   and  the   $XY$  model   belong  to   the   same  universality
class\cite{nho}.  For critical dynamics of a superfluid, however,  
one needs to use the full planar magnet model in which the 
role of the third component of the pseudospin is crucial.\cite{hohenberg}

In  the  pseudospin  notation,  the  planar  magnet  model  takes  the
following form:
\begin{equation}
H=-J\sum_{<ij>}(S_{i}^{x}S_{j}^{x}+S_{i}^{y}S_{j}^{y}),
\end{equation}
where    the    summation    is    over   all    nearest    neighbors,
$\vec{S}_{i}=(S_{i}^{x},S_{i}^{y},S_{i}^{z})$, and $J$ sets the energy
scale. In the usual  $XY$ model the two-component pseudo-magnetization
corresponds to the superfluid order  parameter. In the planar magnet
model, the third  component corresponds to the particle  density
and it is necessary in order to study the dynamics.

In our  calculations, we  use a bar-like  geometry, i.e. a  $H \times
H\times L$ lattice with $L \gg H$. This geometry is chosen in order to
mimic  the  pore  geometry   used  in  experimental  studies.  In  our
calculations,  open  boundary  conditions   are  imposed  in  the  $H$
direction,  and in  the  $L$ direction  we  applied periodic  boundary
conditions.

We use a hybrid Monte  Carlo procedure\cite{krech} which consists of a
combination  of  steps  using   the  Metropolis  update,  the  Cluster
update\cite{wolff},  and  the  over-relaxation  algorithm\cite{brown}.
Using  this hybrid algorithm,  we generate  approximately 3,000-10,000
uncorrelated configurations from the equilibrium canonical ensemble at
a  given  temperature.   Each   configuration  is  evolved  using  the
equations of  motion for  the planar magnet  model which are  given as
follows\cite{krech,hohenberg}
\begin{equation}
\frac{d}{dt}\vec{S}_{i}=\frac{\partial H}{\partial \vec{S}_{i}} \times
\vec{S}_{i}.
\end{equation}
Starting  from a  particular  initial spin  configuration, we  perform
numerical  integration   of  these  equations   of  motion.  Following
Ref.\cite{krech2}   we   use   a  recently   developed   decomposition
method\cite{de} where the integration is carried out to a maximum time
$t_{max}$ (typically of  the order of $t_{max}$=400) with  a time step
$\delta  t$=0.05.  We  made  sure  that this  way  we  determined  the
real-time history  of every  configuration within a  sufficiently long
interval of  time (0$\leq t  \leq t_{max}$).  Finally, we  compute the
thermal average  of a time-dependent  observable (such as  the thermal
current-current correlation function) by averaging over all the values
of  the observable obtained  by evolving  all the  independent initial
equilibrium  configurations  generated  via  the  hybrid  Monte  Carlo
procedure.
 
Compared to calculating  static critical properties, the computation
of dynamical  properties is  far more CPU  time intensive  and demands
large computational  resources.  The computations  described here were
carried out on a dedicated massively parallel cluster of 64 CPUs which
was designed by our group to achieve high performance to cost ratio.

We computed the thermal conductivity on $H \times H\times L$ lattices,
where $H$=6,8,10,12,14,20 and  $L= 5H$. The thermal conductivity
$\lambda$ of liquid  $^{4}He$ at  a given temperature  $T$ can  be calculated
using the dynamic current-current correlation function\cite{krech}:
\begin{equation}
\lambda=\frac{1}{k_{B}T\chi_{zz}}\frac{2}{\pi}\int^{\infty}_{0}dt \sum
_{i} <j^{z}_{0}(0)j^{z}_{i}(t)>,
\end{equation}
where the out-of-plane static susceptibility
\begin{equation}
\chi_{zz}=<M^{z2}>/(k_{B}TL^{3})
\end{equation}
is  needed for  normalization.  The $z$-component  $j^{z}_{i}$ of  the
current density $\vec{j}_{i}$ associated with the lattice point $i$ is
defined by
\begin{equation}
j_{i}^{z}=J(S_{i}^{y}S_{i+e_{z}}^{x}-S_{i}^{x}S_{i+e_{z}}^{y}),
\end{equation}
where  the notation  $i+e_{z}$  denotes the  nearest  neighbor of  the
lattice site $i$ in the $z$ lattice direction.

Now, we would  like to examine the finite-size  scaling hypothesis for
the  thermal resistivity $R(t,H)=1/\lambda(t,H)$,  and to  compare our
results  with  the   existing  experimental  results\cite{kahn}.   The
dependence upon $t$  of the bulk thermal resistivity  can be described
by the power law
\begin{equation}
R(t)=R_{0}t^{\pi},
\label{res1}
\end{equation}
where  $\pi$ is  a  dynamic critical  exponent.   Using Eq.~\ref{res1},  the
finite-size scaling expression for the thermal resistivity $R(t,H)$ is
given by
\begin{equation}
R(t,H)H^{\pi/\nu}=f(tH^{1/\nu}),
\end{equation}
where  the  function  $f(x)$  is  universal and  $\nu$=0.6705  is  the
critical exponent of the correlation length\cite{gold}.

\section{Results}

In  this section,  we  calculate the thermal resistivity, we
examine its scaling behavior with respect to $H$ and then
we compare the scaling function with the experimental results. 
To calculate these observables with small statistical errors
even with our utilization of the most recent numerical advances
and with using the 64-node dedicated cluster, it requires signinificant
amount of time of high-throughput computation. 

\begin{figure}[htp]
\includegraphics[width=\figwidth]{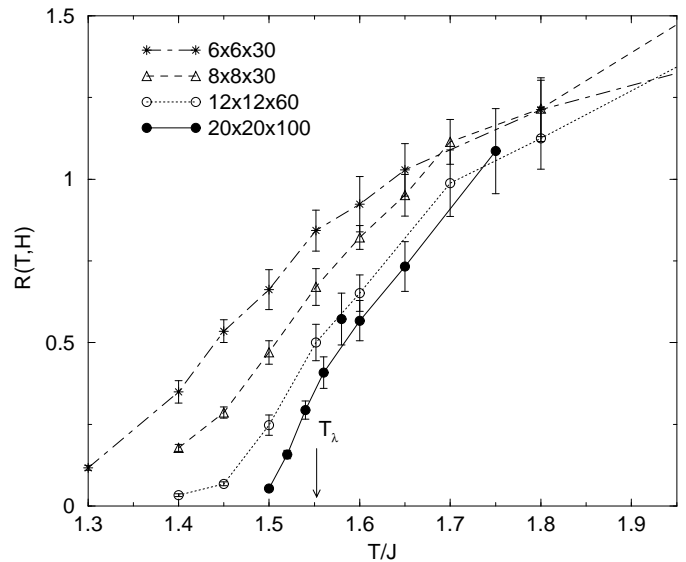}
\caption{Thermal resistivity $R(T,H)$  versus temperatures for bar-like 
lattices with  sizes that correspond to $H=6,8,12,20$ and $L=5 H$. 
The bulk $T_{\lambda}=1.5518$ is also shown.
}
\label{fi-1}
\end{figure}

\begin{figure}[htp]
\includegraphics[width=\figwidth]{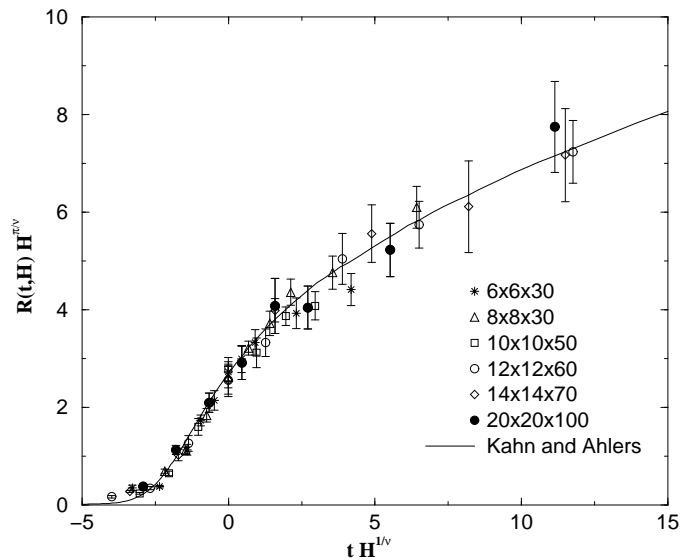}
\caption{The   universal  function   $f(x)$   obtained  for   bar-like
geometry. The solid line  represents the available experimental results
for pore-like  geometry.  In the experimental  results the resistivity
scale and the temperature scale are used as free parameters.}
\label{fi-2}
\end{figure}

\begin{figure}[htp]
\includegraphics[width=\figwidth]{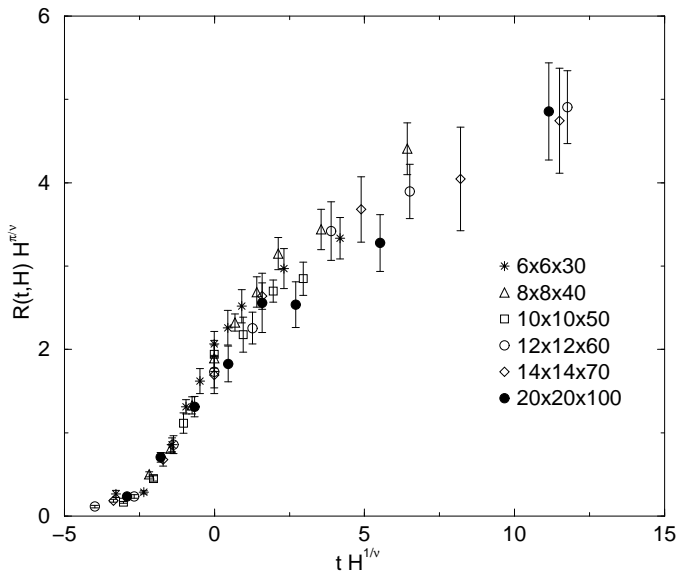}
\caption{The   universal  function   $f(x)$   obtained  for   bar-like
geometry using the theoretical value of $\pi=0.335$. }
\label{fi-2b}
\end{figure}

Fig.~\ref{fi-1} shows some of our results for the thermal resistivity $R(T,H)$
as a function of temperature $T$  for various lattice
sizes with open boundary conditions  in the $H$ direction. Our results
for $R(T,H)$  for several values  of $H$ and  $T$ are given  in Table~
\ref{table1} and ~\ref{table2}.  We wish to make $L$ large enough
so that finite-size effects  due to   $L$  are smaller than  
our  statistical errors.   We  have found  that
taking  $L\approx 5H$ and  applying  periodic boundary  conditions along  the
direction of  $L$ introduces insignificant finite size  effects due to
the finite size of $L$ for the temperature range studied here.   Since
we wish to  remain in the 3D critical region,  it appears that keeping
$L=5 H$ introduces small finite-size effects due to $L$ in this region. 
Lowering the temperature further, when the correlation length 
becomes comparable to
$L$, the value of $R$ is  influenced by the finite size of $L$. In this
region the  system exits the 3D critical region and it behaves as 
a one-dimensional system.  

Notice in  Fig.~\ref{fi-1} that  the thermal  resistivity feels
strong  finite-size effects  due to  the bar  thickness $H$.  The  
arrow shows the bulk transition temperature $T_{\lambda} =1.5518$ 
obtained from Monte Carlo simulation using the planar magnet model\cite{nho}.
In bulk helium  $R(t)$  approaches zero  as  the  bulk transition  temperature
$T_{\lambda}$ is approached from above.

\begin{table}[bth] \centering
 \begin{tabular}{|l|l|l|l|l|l|} \hline $T/J$ &  $H=6$ & $H=8$ & $H=10$
 & $H=12$ & $H=14$ \\ \hline 1.40 & 0.350(34) & 0.178(11) & 0.053(4) &
 0.033(4) & \\  1.45 & 0.535(35) & 0.286(17) &  0.143(10) & 0.068(6) &
 0.050(4) \\  1.50 & 0.662(61) &  0.470(36) & 0.353(38)  & 0.247(31) &
 0.182(21) \\ 1.5518  & 0.843(62) & 0.670(56) &  0.614(53) & 0.501(56)
 &0.452(59) \\  1.60 & 0.923(84)  & 0.821(36) & 0.688(67)  & 0.652(55)
 &0.706(42) \\ 1.65 & 1.028(81) &  0.951(64) & 0.854(42) & & \\ 1.70 &
 & 1.114(69) & 0.901(63) & 0.988 (102)&0.984(105)\\ 1.80 & 1.213(97) &
 1.216(86) & & 1.125(94) &1.081(166) \\ \hline \end{tabular}
\vspace{.2in}
\caption
{Calculated results for the  thermal resistivity for lattices $H\times
H\times  L$  with  $L\approx 5   H$  and  $H=6.8,10,12,14$.   The  number  in
parenthesis gives the error in the last decimal places.}
\label{table1} 
\end{table}

\begin{table}[bth] \centering
 \begin{tabular}{|l|l|} \hline $T/J$ & $H=20$\\ \hline 1.50 & 0.053(3)
 \\ 1.52 & 0.158(13) \\ 1.54 & 0.294(27) \\ 1.56 & 0.408(48) \\ 1.58 &
 0.572(79)  \\  1.60  &  0.567(62)  \\  1.65 &  0.733(76)  \\  1.75  &
 1.086(130)\\ \hline \end{tabular}
\vspace{.2in}   \caption  {   Calculated  results   for   the  thermal
resistivity for an $20\times 20\times 100$ size lattice.}
\label{table2}
\end{table}

We wish to avoid using any adjustable parameters to obtain scaling
of our results. Thus, we need to examine if our results
obey scaling using the known values of the critical 
exponents $\nu$ and $\pi$. The value of $\nu$ is accurately
known from theoretical and experimental studies of static critical 
properties and we shall
use the value $\nu=0.6705$ as determined by Goldner and Ahlers\cite{gold}.
There is less agreement between theory and experiment on the 
actual value of the dynamical critical exponent $\pi$.
Ahlers\cite{ahlers3} used a power law fit to the data of  Tam and
Ahlers\cite{tam}  for   their   ``Cell    $F$''  and he found
the  value $\pi=0.4397$.  However, the  dynamic  scaling   theory\cite{ds}  had
predicted a divergence in $\lambda$  with a critical exponent given by
$\pi=\nu/2\approx 0.335$.  

Fig.~\ref{fi-2}  shows  a  scaling  plot of  the  thermal  resistivity
scaling  function $f(x)=R(t,H)t^{\pi/\nu}$  versus the  scaled reduced
temperature  parameter   $x=t  H^{1\over  \nu}$,   where  the  reduced
temperature  is  taken relative  to  the  bulk transition  temperature
$T_{\lambda}$.  Our Monte Carlo  data collapse onto a universal curve
using  the  value of  $\pi\approx 0.44$  determined by 
Ahlers\cite{ahlers3}.
In Fig.~\ref{fi-2}  we compare our universal function  $f(x)$ with the
experimental data  obtained  by Kahn and  Ahlers\cite{kahn} represented
by a solid line. In
order to do this, we  used two multiplicative constants as free fitting
parameters,  one multiplying  the scale  of $x$  axis and  another the
scale  of  $y$.  The  agreement  between  Monte  Carlo simulation  and
experiment is quite  satisfactory. 
In the past it has been demonstrated\cite{sm,mc5} that
the boundary conditions play a significant role in defining 
the universal function
$f(x)$. We believe  that if we use more  realistic boundary conditions,
such as  Dirichlet boundary conditions, along the  $H$-direction we can
reduce the number of fitting parameters to only one.

Using  the theoretical value of $\pi$,  the  results of  our simulation
also  collapse  on  a  different  scaling  function given in 
Fig.~\ref{fi-2b} . However, if we attempt to fit 
the scaling curve with the experimental resistivity of Kahn and Ahlers 
we obtained a lower quality fit than that of Fig.~\ref{fi-2}.

In  summary we  have calculated  the thermal  resistivity  $R(t,H)$ of
liquid $^{4}He$  in  a pore-like  geometry  (on a $H  \times  H \times  
L$  lattice)
applying  open  boundary  conditions   in  the  $H$  direction.  
We have been able to demonstrate the validity of
finite-size scaling theory and we  obtained the thermal resistivity 
scaling  function $f(x)$ using known values for the critical exponents
and no adjustable parameters. 
In addition, the scaling  function $f(x)$  for   $R(t,H)$ 
agrees rather well with experimental data using the temperature scale
and thermal resistivity scale as free parameters.

\section{acknowledgments}

This  work  was  supported  by  the  National  Aeronautics  and  Space
Administration under grant no. NAG3-1841.

\end{document}